\documentstyle[12pt]{article}
\advance\textheight by \topskip
\begin{document}
\thispagestyle{empty}
\begin{flushright}
SU--ITP--95--5\\
hep-th/9503097\\
March  20, 1995\\
\end{flushright}
\vskip 2 cm
\begin{center}
{\LARGE\bf Inflation  with Variable $\Omega$}
\vskip 1.7cm

 {\bf Andrei Linde}\footnote{
E-mail: linde@physics.stanford.edu},
\vskip 1.5mm
Department of Physics, Stanford University, \\
Stanford, CA 94305--4060, USA
\end{center}
\vskip 2cm

{\centerline{\large\bf Abstract}}
\begin{quotation}
\vskip -0.4cm
 We  propose a simple version of chaotic inflation  which  leads to a
division of the universe into infinitely many open  universes with all possible
values of $\Omega$ from $1$ to $0$.

\end{quotation}
 \newpage

\baselineskip=16pt

Flatness of the universe ($\Omega = 1$) for a long time has been considered as
one of the most definite predictions of the inflationary theory. However,  what
will happen if ten years from now it will be shown beyond any reasonable doubt
that our universe is open or closed? Would it mean that inflationary theory is
wrong?

Apparently, many experts do not think so. So far we do not have any alternative
solution to the homogeneity, isotropy,  horizon and monopole problems.
Inflationary theory provides a natural mechanism for generation of
perturbations necessary for galaxy formation. It would be much better to modify
this scenario so as to produce $\Omega \not = 1$ instead of simply giving up
and living without any consistent cosmological theory at all.

Indeed, it is possible to have $\Omega \not = 1$ in inflationary cosmology. It
is especially simple in the case of a closed universe, $\Omega > 1$. For
example one can consider a particular version of the chaotic inflation scenario
\cite{Chaotic} with the effective potential
\begin{equation}\label{1}
V(\phi) = {m^2 \phi^2\over 2}\, \exp{\Bigl({\phi\over CM_{\rm P}}\Bigr)^2} \ .
\end{equation}
Potentials of a similar type often appear in supergravity. In this theory
inflation occurs only in the interval ${M_{\rm P}\over 2} {\
\lower-1.2pt\vbox{\hbox{\rlap{$<$}\lower5pt\vbox{\hbox{$\sim$}}}}\ } \phi {\
\lower-1.2pt\vbox{\hbox{\rlap{$<$}\lower5pt\vbox{\hbox{$\sim$}}}}\ }
CM_{\rm P}$. The most natural way to realize inflationary scenario in this
theory is
to assume that the universe was created ``from nothing''  with the field $\phi$
in the interval  ${M_{\rm P}\over 2} {\
\lower-1.2pt\vbox{\hbox{\rlap{$<$}\lower5pt\vbox{\hbox{$\sim$}}}}\ } \phi {\
\lower-1.2pt\vbox{\hbox{\rlap{$<$}\lower5pt\vbox{\hbox{$\sim$}}}}\ }
CM_{\rm P}$. According to \cite{Creation}, the probability of this process is
suppressed by
\begin{equation}\label{2}
P \sim \exp\Bigl(-{3M^4_{\rm P}\over 8V(\phi)}\Bigr) \ .
\end{equation}
Therefore the maximum of the probability appears near the upper range of values
of the field $\phi$ for which inflation is possible, i.e. at $\phi_0 \sim C
M_{\rm P}$.
The probability of such an event will be so strongly suppressed that the
universe will be formed almost ideally homogeneous and spherically symmetric.
As  pointed out in   \cite{Lab}, this  solves the homogeneity, isotropy and
horizon problems even before inflation really takes over. Then the size of the
newly born universe in this model expands by the factor $\exp({2\pi
\phi_0^2M_{\rm P}^{-2}})\sim \exp({2\pi C^2})$ during the
stage of inflation \cite{MyBook}. If $C {\
\lower-1.2pt\vbox{\hbox{\rlap{$>$}\lower5pt\vbox{\hbox{$\sim$}}}}\ } 3$, i.e.
if $\phi_0 {\
\lower-1.2pt\vbox{\hbox{\rlap{$>$}\lower5pt\vbox{\hbox{$\sim$}}}}\ } 3M_{\rm P}
\sim 3.6\times 10^{19}$ GeV, the
universe expands  more than $e^{60}$ times, and  it becomes very flat.
Meanwhile, for $C \ll 3$ the universe always remains ``underinflated'' and very
curved, with $\Omega \gg 1$. We emphasize again that in this particular model
``underinflation" does not lead to any problems with homogeneity and isotropy.
The only problem with this model is that in order
to obtain $\Omega$ in the interval between $1$ and $2$ at the present time one
should have the constant $C$ to be fixed somewhere near $C = 3$ with an
accuracy of few percent. This is a fine-tuning, which does not sound very
attractive. However, it is important to realize that we are not talking about
an exponentially good precision; accuracy of few percent is good enough.

It is much more complicated to obtain an open universe, which would be
``underinflated'' and still homogeneous, see e.g. \cite{Tkachev}. However, it
also proves to be possible. The basic idea goes back to the papers by Coleman
and De Luccia \cite{CL} and by Gott \cite{Gott} who pointed out that the space
inside a bubble formed during the false vacuum decay looks like an open
universe. It is amazing that one can hide an infinitely large universe inside
one bubble! (Of course, the trick works only if the bubbles do not collide, but
this condition can be easily satisfied if the probability of bubble formation
is exponentially small.) This mechanism was further explored in the papers by
other authors, notably by Sasaki, Tanaka, Yamamoto and Yokoyama \cite{STYY}.
However, as a result of the tunneling one would obtain an open universe which
would be almost completely empty, with infinitesimally small $\Omega$.

A significant progress in this direction has been achieved recently, when
Bucher, Goldhaber and Turok suggested that the interior of the bubble after
tunneling should continue expanding exponentially, due to additional stage of
inflation, just like in the new or chaotic inflation scenario \cite{BGT}. Then
the first stage of inflation and the spherical symmetry of the created bubble
take care of the large scale homogeneity and isotropy, whereas the total size
of the universe grows after the tunneling due to the second stage. This stage
is largely responsible for density perturbations produced during inflation.
The theory of these density perturbations and the corresponding cosmogony was
developed by several authors even before this particular mechanism was
suggested, but this mechanism adds some new distinctive features to the  large
scale part of the spectrum \cite{Open}.

 Unfortunately, just as in the closed-universe case discussed above, the
realization of this scenario suggested by Bucher, Goldhaber and Turok
\cite{BGT} requires fine tuning. The best model suggested by them recently
\cite{Bucher} was the chaotic inflation scenario  with the potential $V(\phi) =
{m^2\over 2} \phi^2 - {\alpha\over 3} \phi^3 + {\lambda\over 4}\phi^4$
\cite{Kof}. In order to  get an open inflationary  universe in this model   it
was necessary to adjust its parameters  in such a way as to ensure that the
tunneling occurs to the point   $\phi \sim 3 M_{\rm P}$ with the
accuracy of few percent. Also, the tunneling should occur to the part of the
potential which   almost does not change it slope during inflation at smaller
$\phi$, since otherwise one does not   obtain scale-invariant density
perturbations. One of the necessary
conditions was that the barrier should be very narrow. Indeed, if $V'' \ll H^2$
at the barrier, then the tunneling occurs to its  top, as in the Hawking-Moss
case \cite{HM}; see \cite{Lab} for the interpretation of this regime. If this
happens, the large scale density perturbations become huge,
${\delta\rho\over \rho} \sim {H^2\over \dot\phi} > 1$, since $\dot\phi = 0$ at
the maximum. In order to avoid this problem the authors assumed that the field
$\phi$ had a nonminimal kinetic term. Thus the model gradually became not only
fine-tuned, but also rather complicated. It does not discredit the whole idea,
it is nice to have at least one working realization of the scenario outlined in
\cite{BGT}, but with all these complications it becomes very tempting to find a
more natural realization of the chaotic inflation scenario which would give
inflation with $\Omega < 1$.

In this paper we will try to do it. The model is so simple that its description
will take less space than this long introduction. Let us consider a model of
two noninteracting scalar fields, $\phi$ and $\sigma$, with the effective
potential
\begin{equation}\label{3}
V(\phi, \sigma) = {m^2\over 2}\phi^2 + V(\sigma) \ .
\end{equation}
Here $\phi$ is a weakly interacting inflaton field, and $\sigma$, for example,
can be the field responsible for the symmetry breaking in GUTs. We will assume
that $V(\sigma)$ has a minimum at $\sigma = 0$, just as in the old inflationary
theory. The shape of the potential can be, e.g., ${M^2\over 2} \sigma^2 -
{\alpha M } \sigma^3 + {\lambda\over 4}\sigma^4$, but it is not essential;
no fine tuning of the shape of this potential will be required.

Note that so far we did not make any unreasonable complications to the standard
chaotic inflation scenario; at large $\phi$ inflation is driven
by the field $\phi$, and the GUT potential is necessary in the theory anyway.
In order to obtain density perturbations of the necessary amplitude the mass
$m$ of the scalar field $\phi$ should be of the order of $10^{-6} M_{\rm P}
\sim
10^{13}$ GeV \cite{MyBook}.

Inflation begins at $V(\phi, \sigma) \sim M_{\rm P}^4$. At this stage
fluctuations of
both fields are very strong, and the universe enters the stage of
self-reproduction, which finishes (for the field $\phi$) only when it becomes
smaller than $M_{\rm P} \sqrt{M_{\rm P}\over m}$ and the energy density drops
down to $m
M_{\rm P}^3  \sim 10^{-6} M_{\rm P}^4$ \cite{MyBook}. Quantum fluctuations of
the field
$\sigma$ in some parts of the universe put it directly to the minimum of
$V(\sigma)$, but in some other parts the scalar field $\sigma$ appears in the
local minimum of $V(\sigma)$ at $\sigma  = 0$. Since the energy density in such
domains will be greater, their volume will grow with the greater speed, and
therefore they will be especially important for us. One may worry that all
domains withe $\sigma = 0$
will tunnel to the minimum of $V(\sigma)$ at the stage when the field $\phi$
was very large and quantum fluctuations of the both fields were large too.
However, this decay can be easily suppressed if one introduces a
small interaction $g^2\phi^2\sigma^2$ between these two fields, which
stabilized the state with $\sigma = 0$ at large $\phi$.
 Note that even when the
field $\phi$ drops down to $\phi = 0$, inflation in the domains with $\sigma =
0$ continues, being supported by the false vacuum energy $V(\sigma = 0)$, until
the field $\sigma$ tunnels to the  minimum of $V(\sigma)$.

Here comes the main idea of our scenario. Because the fields $\sigma$ and
$\phi$ do not interact with each other, and the dependence of the probability
of tunneling on the vacuum energy at the GUT scale is negligibly small
\cite{CL}, tunneling to the minimum of $V(\sigma)$ may occur with equal
probability at all sufficiently small values of the field $\phi$. The
parameters of the bubbles of the field $\sigma$ are determined by the mass
scale corresponding to the effective potential $V(\sigma)$. This mass scale in
our model is much greater than $m$. Thus the duration of tunneling in the
Euclidean ``time'' is much smaller than $m^{-1}$. Therefore the field $\phi$
practically does not change its value during the tunneling.  Note, that if
the probability of decay at a given $\phi$ is small enough, then it does not
destroy the whole vacuum state $\sigma = 0$ \cite{GW}; the bubbles of the new
phase are produced all the way when    the field $\phi$ rolls down to $\phi =
0$. In this process  the universe  becomes filled with
(nonoverlapping) bubbles immersed in the false vacuum state with $\sigma = 0$.
Interior of each of these bubbles   represents an open universe. However, these
bubbles   contain {\it different} values of the field $\phi$, depending on the
value of this field at the  moment when the bubble formation occurred. If the
field $\phi$ inside a bubble is smaller than $3 M_{\rm P}$, then the universe
inside
this bubble will have a vanishingly small $\Omega$, at the age $10^{10}$ years
after the end of inflation it will be practically empty, and life of our type
could not exist there.  If the field $\phi$ is much greater than $3 M_{\rm P}$,
the
universe inside the bubble will be almost exactly flat, $\Omega = 1$, as in the
simplest version of the chaotic inflation scenario. It is important, however,
that {\it in  an eternally existing self-reproducing universe there will be
infinitely many universes containing any particular value of $\Omega$, from
$\Omega = 0$ to $\Omega = 1$}.

Of course, one can argue that we did not solve the problem of fine tuning, we
just transformed it into the fact that only a very small percentage of all
universes will have, say,  $0.2 <\Omega < 0.3$. However, first of all, we
achieved our goal in a very simple theory, which does not require any
artificial potential bending and nonminimal kinetic terms. Then, there may be
some reasons why it is preferable for us to live in a universe with a small
(but not vanishingly small) $\Omega$. Indeed, the total volume of the bubbles
with $\Omega = 1$ grows at a much smaller rate after the phase transition.
Thus, the later the phase transition happen, the more volume we get. This
emphasizes the universes with small $\Omega$. On the other hand, we cannot live
in empty universes when $\Omega$ is too small. The percentage of the universes
with different $\Omega$ can be strongly influenced by introducing a small
coupling $g^2\phi^2\sigma^2$, which stabilizes the state $\sigma = 0$ for large
$\phi$. The tunneling becomes possible only for sufficiently small $\phi$. This
suppresses the number of bubbles with $\Omega = 1$.

We do not want to pursue this line of arguments any further. Comparison of
volumes of different universes  in the context of a theory of a
self-reproducing inflationary universe  is a very ambiguous task, since in this
case we must compare infinities \cite{LLM}. If we would know how to solve  the
problem of measure in quantum cosmology, perhaps we would be able to obtain
something similar to an open universe without any first order phase transitions
\cite{OPEN}.  In the meantime, it is already encouraging that in our scenario
there are infinitely many inflationary universes with any particular value of
$\Omega$. It may happen that  the only way to find out whether we live in one
of them is to make observations.

Some words of caution are in order here. The bubbles produced in our scenario
are not {\it exactly} open universes. Indeed, in the models discussed in
\cite{CL}--\cite{BGT} the time of reheating (and the temperature of the
universe after the reheating) was synchronized with the value of the scalar
field inside the bubble. In our case the situation is very similar, but not
exactly \cite{LM}. Suppose that the Hubble constant induced by $V(0)$ is much
greater than the Hubble constant related to the energy density of the scalar
field $\phi$. Then the speed of rolling of the scalar field $\phi$ sharply
increases inside the bubble. Thus, in our case the field $\sigma$ synchronizes
the motion of the field $\phi$, and then the hypersurface of a constant field
$\phi$ determines the hypersurface of a constant temperature. In the models
where the rolling of the field $\phi$ can occur only inside the bubble (we will
discuss such a model shortly)  the  synchronization is precise, and everything
goes as in the models of refs. \cite{CL}--\cite{BGT}. However, in our simple
model the scalar field $\phi$ moves down outside the bubble as well, even
though it does it very slowly. Thus,   synchronization of   motion of the
fields $\sigma$ and $\phi$  is not precise; hypersurface of a constant $\sigma$
ceases to be a hypersurface of a constant density. For example, suppose that
the field $\phi$ has taken some value $\phi_0$ near the bubble wall when the
bubble was just formed. Then the bubble expands, and during this time the field
$\phi$ outside the wall  decreases, as $\exp \Bigl(-{m^2t\over 3 H(0)}\Bigr)$,
where $H(0) = \sqrt{8\pi V(0)\over 3 M_{\rm P}^2}$ \cite{MyBook}. At the moment
when the bubble expands $e^{60}$ times, the field $\phi$ in the region just
reached by  the bubble wall decreases to  $\phi_o\exp \Bigl(-{20 m^2\over
H^2(0)}\Bigr)$ from its original value $\phi_0$. The universe inside the bubble
is a homogeneous open universe only if this change is negligibly small. This
may not be a real problem. Indeed,  let us assume that $V(0) = M^4$, where $M =
10^{17}$ GeV. In this case $H(0) = 1.7 \times 10^{15}$ GeV, and for $m =
10^{13}$ GeV one obtains ${20 m^2\over   H^2(0)} \sim 10^{-4}$. In such a case
a typical degree of distortion of the picture of a homogeneous open universe is
very small.

Still this issue deserves careful investigation. When the bubble wall continues
expanding even further, the scalar field outside of it eventually drops down to
zero. Then there will be no new matter created near the wall.  Instead of
infinitely large homogeneous open universes we are obtaining   spherically
symmetric islands of a size much greater than the size of the observable part
of our universe. We do not know whether this unusual picture is a curse or a
blessing for our model. Is it possible to consider different parts of the same
exponentially large island as domains of different ``effective'' $\Omega$? Can
we attribute some part of the dipole anisotropy of the microwave background
radiation to the possibility that we live somewhere outside of the center of
such island?

Another potential problem associated with this model is the possibility that
the density perturbations on the horizon scale can appear larger than expected.
Indeed, the Hubble constant before the tunneling in our model was much greater
than the Hubble constant after the tunneling. This may lead to very large
density perturbations on the scale comparable to the size of the bubble. Again,
this may not be a real problem, since in the new coordinate system, in which
the interior of the bubble looks like an  open universe, the distance from us
to the bubble walls is infinite. However,   to be on a safe side it would be
nice to have a model where we do not have any problems with synchronization and
with the large jumps of the Hubble constant. This can be achieved by a
generalization (simplification) of our model (\ref{3}):
\begin{equation}\label{4}
V(\phi,\sigma) = {g^2\over 2}\phi^2\sigma^2 + V(\sigma) \ .
\end{equation}
We eliminated the massive term of the field $\phi$ and added explicitly the
interaction ${g^2\over 2}\phi^2\sigma^2$, which, as we have mentioned already,
is desirable for stabilization of the state $\sigma = 0$ at large $\phi$. Note
that in this model the line $\sigma = 0$ is a flat direction in the
($\phi,\sigma$) plane. At large $\phi$ the only minimum of the effective
potential with respect to $\sigma$ is at the line $\sigma = 0$.  To give a
particular example, one can take $V(\sigma) = {M^2\over 2} \sigma^2 -{\alpha M
} \sigma^3 + {\lambda\over 4}\sigma^4 +V_0$. Here $V_0$ is a constant which is
added to ensure that $V(\phi,\sigma) = 0$ at the absolute minimum of
$V(\phi,\sigma)$.  In this case the minimum of the potential $V(\phi,\sigma)$
at $\sigma \not = 0$ is deeper than the minimum at $\sigma = 0$ only for $\phi
< \phi_c$, where $\phi_c = {M\over g}\sqrt{{2\alpha^2\over  \lambda} -1}$. This
minimum for $\phi = \phi_c$ appears at $\sigma = \sigma_c = {2\alpha M\over
\lambda}$.

The bubble formation becomes possible only for $\phi < \phi_c$. After the
tunneling the field $\phi$ acquires an effective mass $m = g\sigma$ and begins
to move towards $\phi = 0$, which provides the mechanism for the second stage
of inflation inside the bubble. In this scenario evolution of the scalar field
$\phi$ is exactly synchronized with the evolution of the field $\sigma$, and
the universe inside the bubble appears to be open.

Effective mass of the   field $\phi$ at the minimum of $V(\phi,\sigma)$ with
$\phi = \phi_c$, $\sigma = \sigma_c = {2\alpha M\over  \lambda}$ is   $m =
g\sigma_c = {2g\alpha M\over  \lambda}$. With a decrease of the field $\phi$
its effective mass at the minimum of $V(\phi,\sigma)$ will grow, but not
significantly. Let us consider, e.g., the theory with  $\lambda =  \alpha^2 =
10^{-2}$, and $M \sim 5\times10^{15}$ GeV, which seems quite natural.  In this
case it can be shown that the effective mass $m$ is equal to $2gM$ at $\phi =
\phi_c$, and then it grows by only $25\%$ when the field $\phi$ changes all the
way down from  $\phi_c$ to $\phi = 0$.   As we already mentioned, in order to
obtain the proper amplitude of density perturbations  one should have $m \sim
10^{13}$ GeV.   In our case $m \sim 10^{13}$ GeV   for  $g \sim 10^{-4}$, which
gives $\phi_c \sim 5\times 10^{19}\  \mbox{GeV} \sim 4M_{\rm P}$.  The bubble
formation becomes possible only for $\phi < \phi_c$. If it happens in the
interval $4M_{\rm P} > \phi > 3 M_{\rm P}$, we obtain a flat universe. If it
appears at $\phi < 3M_{\rm P}$, we obtain an open universe. Depending on the
initial value of the field $\phi$, we can obtain all possible values of
$\Omega$, from $\Omega = 1$ to $\Omega = 0$. The value of the Hubble constant
at the minimum with $\sigma \not = 0$ at $\phi = 3M_{\rm P}$ in our model does
not differ much from the value of the Hubble constant before the bubble
formation. Therefore we do not expect any specific problems with the large
scale density perturbations in this model.
 Note also that the probability of tunneling at large $\phi$ is very small
since the depth of the minimum at $\phi \sim \phi_c$, $\sigma \sim \sigma_c$
does not differ much from the depth of the minimum at $\sigma = 0$. Therefore
the number of flat universes produced by this mechanism will be strongly
suppressed as compared with the number of open universes. Meanwhile, life of
our type is impossible in empty universes with $\Omega \ll 1$. This may provide
us with a tentative explanation of the small value of $\Omega$ in the context
of our model.

As we have seen, the models which can give rise to an open inflationary
universe are very simple, but they lead to a rather complicated dynamics.
Therefore they deserve thorough investigation. The main purpose of this paper
was to show that there exists a wide class of models which can describe an open
inflationary universe. It is still necessary to find out which of these models
could describe observational data in a better way. Note, that there is no need
to have an extremely small coupling constant $\lambda \sim 10^{-13}$ in our
model. Instead of it we have a small coupling $g = 10^{-4}$. Thus at least in
this aspect we are {\it reducing} the level of fine tuning required in the
simplest inflationary models with $\Omega = 1$.    We will return to the
discussion of these and some other  models with $\Omega \not = 1$ in   the
forthcoming publication \cite{LM}.

We would like to conclude this article with some general remarks.  Fifteen
years ago many different cosmological models (HDM, CDM, $\Omega = 1$, $\Omega
\ll 1$, etc.) could describe all observational data reasonably well. The main
criterion for a good theory was its beauty and naturalness. Right now it
becomes more and more complicated to explain  all observational data. In such a
situation cosmologists should remember that the standard theory of electroweak
interactions contains about twenty free parameters which so far did not find an
adequate theoretical explanation. Some of these parameters may  appear rather
unnatural. The best example is the coupling constant of the electron to the
Higgs field,
which is $2\times 10^{-6}$. It is a pretty unnatural number which is fine-tuned
in such a way as to make the electron 2000 lighter than the proton. We do not
have any reason to expect that the cosmological theory will be simpler than
that. It is important, however, that the electroweak theory is based on two
fundamental principles: gauge invariance and spontaneous symmetry breaking. As
far as these principles hold, we can adjust our parameters and wait until they
get their interpretation in a context of a more general theory. It seems that
inflation provides a very good guiding principle for constructing  internally
consistent cosmological models. The simplest versions of inflationary theory
predict a  universe with $\Omega  = 1$. However, it is very encouraging that
this theory, if needed, can be  versatile enough to include models with all
possible values of $\Omega$, without forcing us to give up all  advantages of
inflationary
cosmology. This is an important point which often escapes attention of those
who tries to compare predictions of inflationary cosmology with observational
data. At the present moment  inflation is the only mechanism known to us that
could produce a large homogeneous universe with $\Omega \not = 1$.

The author  is very grateful to  M. Bucher, J. Garc\'{\i}a--Bellido,  L.
Kofman,  A. Mezhlumian, and I. Tkachev for many enlightening discussions.  This
work was
supported in part  by NSF grant PHY-8612280.

%\newpage

\end{document}